\documentclass[aps,twocolumn]{revtex4}
\usepackage[dvips]{graphics,graphicx}
\usepackage{amsmath}
\begin{document}

\title{Probing Quantum Chaos in many-body quantum systems by the induced dissipation}

\author{Anna~A.~Bychek$^1$}
\author{Pavel~S.~Muraev$^2$}
\author{Andrey~R.~Kolovsky$^{1,2}$}
\affiliation{$^1$Kirensky Institute of Physics, 660036 Krasnoyarsk, Russia}
\affiliation{$^2$Siberian Federal University, 660041 Krasnoyarsk, Russia}
\date{\today}
\begin{abstract}
We theoretically analyze the depletion dynamics of an ensemble of cold atoms in a quasi one-dimensional optical lattice where atoms in one of the lattice sites are subject to decay. Unlike the previous studies of this problem in R. Labouvie, {\em et. al}, Phys. Rev. Lett. {\bf 116}, 235302 (2016) we focus on the case where the system is brought to the chaotic regime,  that crucially modifies the depletion dynamics as compared to the regular case.  It is shown that depletion of the affected site results in gradual depletion of the neighbouring sites according to $t^{1/3}$ scaling law. We also show that by measuring occupations of the lattice sites one can extract important information on chaotic dynamics of the original conservative system.
\end{abstract}
\maketitle

{\em 1.}
The field of Quantum Chaos \cite{Gian91,Stoe99,Haak10,Ales16} has emerged in physics in late 70th and early 80th and then rapidly evolved into several branches with one of them addressing non-integrable many-body systems \cite{Mont93,Jacq97,Berm01,66,Elsa15,103}, where the ultimate aim is foundation of the equilibrium and non-equilibrium Statistical Mechanics from the first principles \cite{Elsa15}. The other recent development in physics is the open many-body systems \cite{Witt08,Pros10,Barm11,Baro13}.  In particular, the authors of the cited in the abstract laboratory experiment \cite{Labo16} study dynamics of a Bose-Einstein condensate (BEC) of cold atoms in a quasi one-dimensional optical lattice where one of the lattice sites is constantly depleted by using a tightly focused electron beam. Merging these two developments one comes to the problem of {\em open many-body chaotic systems},  where we may expect some universal dynamics \cite{preprint}.

In the present work we theoretically analyze the system studied in Ref.~\cite{Labo16} yet for the principally different initial condition where the BEC of atoms is brought to the edge of the Brilloine zone. Experimentally this is done by accelerating the lattice or by applying a gradient of the magnetic field for one-half of the Bloch period. As known, at the zone edge the BEC of repulsively interacting atoms exhibits dynamical or modulation instability that indicates the onset of chaos. We are interested in the depletion dynamics which is shown to provide important information about chaotic properties of the closed system. Thus, one can use the induced dissipation as a tool for probing Quantum Chaos in the many-body systems.  

{\em 2.}
The system dynamics is governed by the master equation, 
\begin{equation}
\label{a1}
\frac{d {\cal R}}{dt}=-i[\widehat{H},{\cal R}] - \widehat{{\cal L}}_\gamma({\cal R}) \;,
\end{equation}
on the density matrix  ${\cal R}(t)$ of the ensemble of interacting atoms in a lattice,
\begin{equation}
\label{a2}
\widehat{H}=\omega \sum_{l=1}^L  \hat{n}_l   
  -\frac{J}{2} \sum_{l=1}^L \left( \hat{a}^\dag_{l+1}\hat{a}_l +h.c.\right)
  +\frac{U}{2}\sum_{l=1}^L \hat{n}_l(\hat{n}_l-1) \;,
\end{equation}
where the Lindblad operator $\widehat{{\cal L}}_\gamma({\cal R})$,
\begin{equation}
\label{a3}
\widehat{{\cal L}}_\gamma({\cal R})=\frac{\gamma}{2}
 (\hat{a}^\dagger_l\hat{a}_l{\cal R}-2\hat{a}_l{\cal R}\hat{a}^\dagger_l  + {\cal R}\hat{a}^\dagger_l\hat{a}_l) \;,
\end{equation}
acts only on the single site with the index $l=L/2$. In Eqs.~(\ref{a2}-\ref{a3}) $\omega$ is the frequency of zero oscillations, $J$ the hopping matrix elements, $U$ the microscopic interaction constant, and $\gamma$ the depletion rate. An additional parameter of the system is the mean occupation number of the lattice sites $\bar{n}=N/L$. In the experiment \cite{Labo16} $\bar{n}\approx 700$, $J\approx 230 Hz$, and the macroscopic (mean field) interaction constant $g=U\bar{n}\approx1400Hz$.

Below we solve Eq.~(\ref{a1}) by using the pseudoclassical approach which is based on the notion of the truncated Wigner or Husimi functions \cite{Stee98,Trim08a}. With respect to cold atoms in optical lattices this approach was used, in particular, to analyze Bloch oscillations of interacting atoms. It was  demonstrated in Ref.~\cite{80} that the pseudoclassical approach is capable to reproduce extremely well both results of the exact quantum simulations \cite{80} and the experimental results  \cite{Mein13}.  In the next paragraph we briefly review the main statements of this approach (for more details see Refs.~\cite{103,107}).

{\em 3.}
In the framework of the pseudoclassical  approach the dynamics is described by the distribution function $f({\bf a},t)$ which is a function of time and $L$ complex variables $a_l$, $l=1,\ldots,L$.  Assuming for the moment $\gamma=0$ and neglecting the terms $O(1/\bar{n})$  it satisfies  the Liouville equation 
\begin{equation}
\label{b1}
\frac{\partial f}{\partial t}=\{H,f\}  \;,
\end{equation}
where $\{\ldots,\ldots\}$ denotes the Poisson brackets and $H$ is the Hamiltonian of the classical Bose-Hubbard model,
\begin{equation}
\label{b2}
H=\omega \sum_{l=1}^L  a^*_l a_l -\frac{J}{2}\sum_{l=1}^L (a^*_{l+1}a_l + c.c.) + \frac{g}{2}\sum_{l=1}^L |a_l|^4  \;, 
\end{equation}
where $g=\frac{UN}{L}=U\bar{n}$. (Notice that in Eq.~(\ref{b2}) we apply normalization $\sum_{l=1}^L |a_l |^2=L$.) Commonly, one solves Eq.~(\ref{b1}) by solving the Hamilton equations of motion,
\begin{equation}
\label{b3}
i\dot{a}_l=\frac{\partial H}{\partial a^*_l} = \omega a_l-\frac{J}{2}\left(a_{l+1} +a_{l-1}\right) + g|a_l|^2 a_l  \;,
\end{equation}
and averaging the result over an ensemble of initial conditions with the distribution function $f({\bf a},t=0)$. For example, for relative occupations of the lattice sites $n_l(t)=N_l(t)/\bar{n}$ (which are the quantities measured in the laboratory experiment) we have
\begin{equation}
\label{b0}
n_l(t)={\rm Tr}[\hat{a}_l^\dagger \hat{a}_l {\cal R}(t)]/\bar{n}=\overline{|a_l(t) |^2} \;,
\end{equation}
where the overline denotes the ensemble average.  Since $f({\bf a},t=0)$ is uniquely determined by the initial many-body wave function of the quantum system (\ref{a2}) we refer to this ensemble of initial conditions as the quantum ensemble. Typically, it is a difficult numerical problem to generate the quantum ensemble. Fortunately, for some important many-body states like, for example, the BEC or Mott-insulator states,  quantum ensembles are known explicitly \cite{103,80}.
\begin{figure}
\includegraphics[width=7.5cm,clip]{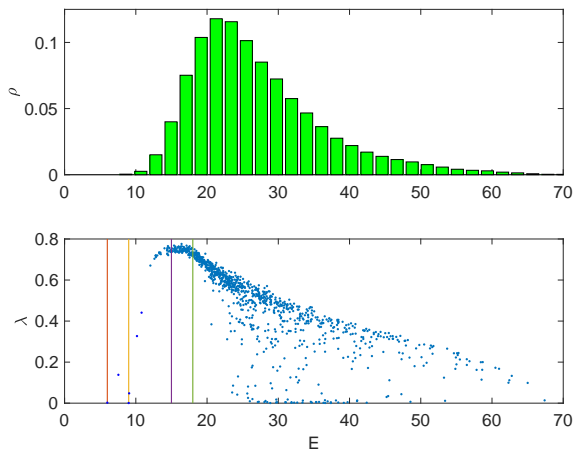}
\caption{The classical Bose-Hubbard model. Upper panel: Relative volume of the energy shell as the function of the shell energy $E$. Lower panel: Lyapunov exponent $\lambda$ of 1000 trajectories with the initial conditions uniformly distributed over the whole phase space. Vertical lines mark energies of the periodic trajectories (\ref{b6}). Parameters are $L=6$, $J=1$, and $g=4$.}
\label{fig1}
\end{figure} 

Next we discuss the classical  Bose-Hubbard model (\ref{b2}). Since it can be viewed as the system of $L$ coupled nonlinear oscillator $h_l$, 
\begin{equation}
\label{b5}
h_l=\omega I_l +\frac{g}{2} I_l^2 \;,\quad I_l=|a_l |^2 \;,
\end{equation}
one expects that its dynamics is generally chaotic. As an example, we consider the case $L=6$ which already captures  the main properties of  larger systems $L\gg 1$. The upper panel in Fig.~\ref{fig1} shows the volume of the energy shell as the function of the shell energy $E$ for $g/J=4$. (We note, in passing, that the depicted histogram reproduces the density of states of the quantum  Bose-Hubbard model \cite{107}.) The lower  panel in Fig.~\ref{fig1} shows the Lyapunov exponent $\lambda$ of different  trajectories ${\bf a}(t)$ with the initial conditions uniformly distributed over the whole phase space, which is a hyper-sphere defined by the condition  $\sum_{l=1}^L |a_l |^2=L$. Additional vertical lines in Fig.~\ref{fig1}(b) mark the energies of the nonlinear Bloch waves, 
\begin{equation}
\label{b6}
a_l(t)=\exp[i\kappa l+iJ\cos(\kappa) t - igt ] \;,\quad \kappa=2\pi k/L \;,
\end{equation}
which are stable ($|\kappa|<\pi/2$) or unstable  ($|\kappa|>\pi/2$) periodic trajectories of the system. As expected, we find regular trajectories (i.e., vanishing Lyapunov exponent) only for low- and high-energy trajectories while trajectories in the middle of the `spectrum' are chaotic with probability close to unity. 

Chaotic dynamics of the system for $|\kappa|>\pi/2$ implies that the time behaviour of fast variables is a random process. In particular, we consider the variable $\xi(t)$, 
\begin{equation}
\label{b7}
\xi(t)=a_{l+1}(t)+a_{l-1}(t) \;,
\end{equation}
which is the driving force for the $l$th oscillator Eq.~(\ref{b5}). We found that the auto-correlation function of $\xi(t)$ is well approximated by the exponential function, 
\begin{equation}
\label{b8}
\overline{\xi(t)\xi^*(t')} \approx A\exp(-|t-t'|/\tau) \;, 
\end{equation}
\begin{displaymath}
A=\overline{|a_{l+1}(t)|^2}+\overline{|a_{l-1}(t)|^2}=2 \;,
\end{displaymath}
where the correlation time $\tau$ is determined by the Lyapunov exponent $\lambda$ which, in turn, is determined by the ratio $g/J$. Thus, we have some freedom in varying the correlation time $\tau$ by varying the ratio $g/J$.

\begin{figure}
\includegraphics[width=8.5cm,clip]{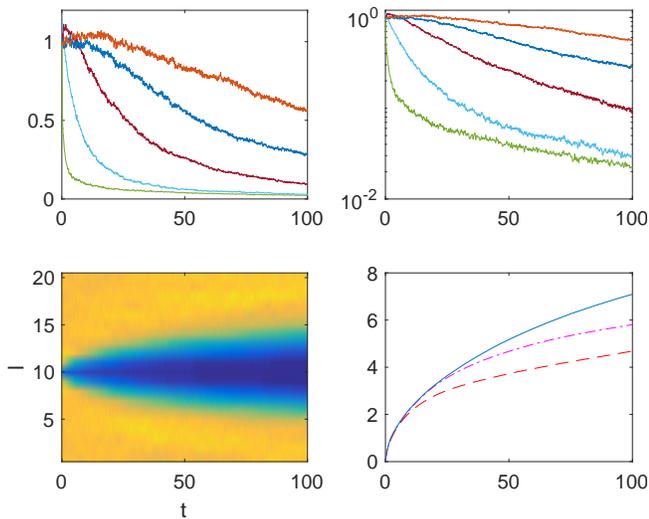}
\caption{Occupations of the lattice sites in the course of time. Upper panels: Occupations of the lattice sites with $0 \le |l| \le L/2+4$ as functions of time in the linear and logarithmic scales. Lower-left panel: Occupation dynamics as a colour map (dark-blue $=0$, bright-yellow $=1$). Lower-right panel: Total number of the depleted particles normalised to $\bar{n}$ for different positions of the weak link $\epsilon J$, see text. Parameters are $L=20$ (periodic boundary conditions), $J=1$, $g=4$, $\gamma=0.1$, and $\epsilon=1/\sqrt{10}$. Average over 1000 trajectories.}
\label{fig3}
\end{figure}

{\em 4.}
Assume now that $\gamma\ne0$. In this case Eq.~(\ref{b1}) should be complimented by the relaxation term ${\cal L}_\gamma(f)$. Again neglecting the terms $O(1/\bar{n})$ we obtain from Eq.~(\ref{a3}) 
\begin{equation}
\label{c1}
{\cal L}_\gamma(f)=\gamma f +\frac{\gamma}{2}\left( a\frac{\partial f}{\partial a} + a^*\frac{\partial f}{\partial a^*} \right) \;,
\end{equation}
where we omit sub-index $L/2$ not to overburden the equation (see Appendix A).  It is also easy to show that this relaxation term modifies Eq.~(\ref{b3}) as
\begin{equation}
\label{c2}
i\dot{a}_l= (\omega-i\gamma\delta_{l,L/2}) a_l - \frac{J}{2}\left(a_{l+1} +a_{l-1}\right) + g|a_l|^2 a_l  \;.
\end{equation}
We run Eq.~(\ref{c2}) for the initial conditions taken from the quantum ensemble for the BEC of atoms accelerated to the edge of the Brillouin zone. Approximately it corresponds to $a_l(t=0)\approx(-1)^l$ with tiny fluctuations of the amplitude and phase that are proportional to $\bar{n}^{-1/2}$ \cite{80}. However, due to the positive Lyapunov exponent [see Fig.~\ref{fig1}(b)] these tiny fluctuations result in completely different trajectories and, hence, only the average over the ensemble has a physical meaning. This average is depicted in Fig.~\ref{fig3} which shows the occupations of the lattice sites and the total number of depleted atoms,
\begin{equation}
\label{c7}
N(t)=\bar{n}\sum_{l=1}^L[1-n_l(t)] \;, 
\end{equation}
as the functions of time. Below we quantify the observed occupation dynamics by using some simple approximations. 

{\em 5.} 
First we address the short-time dynamics ($t<10$ in Fig.~\ref{fig3}) of the central oscillator. Let us for the moment suppress the back action of this oscillator on the other oscillators by changing the hopping between the central well and the neighbouring wells with $l=L/2\pm 1$ from $J$ to a small value $\epsilon J$, $\epsilon\ll 1$. This reduces the amplitude of the stochastic force proportionally to $\epsilon$ and,  simultaneously, divides the whole system into the system of  interest (the central oscillator) and the reservoir (remaining oscillators).  Then the dynamics of the central oscillator is governed by the stochastic equation
\begin{equation}
\label{c3}
i\dot{a}= (\omega-i\gamma) a + g|a|^2 a - \frac{\epsilon J}{2}\xi(t)  \;,
\end{equation}
where, as before, we drop the subindex $l=L/2$. From Eq.~(\ref{c3}) we obtain the equation on the distribution function $f(a,a^*,t)$,
\begin{equation}
\label{c4}
\frac{\partial f}{\partial t}=\{h,f\} + {\cal L}_\gamma(f) +D\frac{\partial^2 f}{\partial a \partial a^*} \;,
\end{equation}
where the relaxation term ${\cal L}_\gamma(f)$ is defined in Eq.~(\ref{c1}) and the diffusion coefficient $D=\epsilon^2 J^2 \tau/2$. (Quantum counterpart of the introduced diffusion term is discussed in Appendix B.) It is easy to prove that the stationary solution of Eq.~(\ref{c4}) is given by the two-dimensional Gaussian
\begin{equation}
\label{c5}
f(a,a^*)=\frac{1}{2\pi \sigma^2}\exp\left(-\frac{|a|^2}{2\sigma^2} \right)  \;,
\end{equation}
where $\sigma^2=D/\gamma$. This determines  the relative stationary occupation of the central site as
\begin{equation}
\label{c6}
\tilde{n}=\frac{\epsilon^2 J^2\tau}{2\gamma} \ll  1\;. 
\end{equation}

Let us now come back to the original system where the artificial parameter $\epsilon=1$.  As seen in Fig.~\ref{fig3}(a-c),  the back action of the central oscillator on the neighbouring oscillators results in gradual decay of the latter. A direct consequence of this is that the stationary value $\tilde{n}$ becomes quasi-stationary,
\begin{equation}
\label{c8}
\tilde{n}_{L/2}(t)=\frac{J^2 n_{L/2\pm 1}(t)\tau}{2\gamma} \;, 
\end{equation}
where $n_{L/2\pm 1}(t)$ is the mean occupation of the nearest sites and it is implicitly assumed that it is still close to unity. Thus, by measuring  the site occupations one can find the correlation time $\tau$ in Eq.~(\ref{b8}).

{\em 6.}
The above results on the short-time dynamics suffice to qualitatively describe the long-time dynamics of the system.  It is a sequence of  step by step oscillator decay starting from the central oscillator. Furthermore, the decay  of every oscillator follows two stages -- first it decays to a quasi-stationary state characterised by some equilibrium value  $\tilde{n}$, which slowly decreases during the second stage, see Fig.~\ref{fig3}(b). (We also mention that by a proper rescaling of the time axis the different curves in Fig.~\ref{fig3}(a,b) can be brought above each other.)

The other approach  to describe the long-time dynamics is the artificial devision of the whole system into the system of interest and the reservoir. In fact, by putting the weak link $\epsilon J$ far enough from the central site we reduce the problem to the known problem of atomic current in the Bose-Hubbard chain, where the first site of the chain is connected to a particle source and the last site to a particle sink (see the recent work \cite{112} and references therein). The dashed, dash-dotted, and solid lines in Fig.~\ref{fig3}(d) show the relative number of depleted particles in the cases where the weak links are located at 4, 6, and 10 sites away from the central well, respectively. Notice the asymptotic linear growth of $N(t)$ in the first two cases which corresponds to a steady state regime with the stationary current. According to Ref.~\cite{112}, the relaxation time to this steady state scales as the chain length to the third power. Inverting this relation we conclude that the number of depleted wells grow proportionally to $t^{1/3}$.

{\em 7.}
We analyzed the depletion dynamics of cold atoms in a quasi one-dimensional optical lattice where atoms in one of the lattice sites  are subject to decay. Experimentally this is done by ionizing the atoms by an electron beam focused on that site \cite{Labo16}. Unlike Ref.~\cite{Labo16},  in the present work we consider the principally different initial state where the BEC of atoms is brought to the edge of the Brilloine zone. In this case the system is chaotic in the sense of both classical and Quantum  Chaos \cite{103}, that crucially modifies the depletion dynamics as compared to the regular case where the BEC is in the ground state (the centre of the Brilloune zone). It is predicted that in the chaotic case depletion of the affected site results in a gradual depletion of the neighbouring sites, so that the total number of the depleted sites grows $\sim t^{1/3}$. We also show that by measuring occupations of the lattice sites one can extract the decay time $\tau$ for correlation functions of the type $C(\tau)=\langle \hat{a}^\dagger_l(t) \hat{a}_l(t+\tau)\rangle$ [here $\hat{a}^\dagger_l(t)$ and $\hat{a}_l(t)$ are the creation and annihilation operators in the Heisenberg representation]. Keeping in mind  that classically the correlation time $\tau$ is determined by the Lyapunov exponent $\lambda$ of the classical Bose-Hubbard model, the proposed experimental studies of the depletion dynamics will shed light on the long-standing question on the meaning of the classical Lyapunov exponent in the quantum realm. 

This work has been supported through Russian Science Foundation grant N19-12-00167.  The authors are grateful to D.~N.~Maksimov for stimulating discussions.


\newpage
\section{Appendix A}

In the pseudoclassical approach operators are given by their Wigner-Weyl images. For example, the operator $\hat{a}/\sqrt{\bar{n}}$ corresponds to $a=(q+ip)/\sqrt{2}$ and the operator $\hat{a}^\dagger/\sqrt{\bar{n}}$ to $a^*=(q-ip)/\sqrt{2}$.  (Notice that in the pseudoclassical approach one works with the creation and annihilation operators which commute to the effective Planck constant $\hbar'=1/\bar{n}$.) Knowing images of two arbitrary operators $A$ and $B$  the image of their product  is given by the equation \cite{Grae18}
\begin{equation}
\label{f1}
A\star B=A\exp\left[ \frac{\hbar'}{2i} \left( 
\frac{\partial^\leftarrow}{\partial q}\frac{\partial^\rightarrow}{\partial p}
-\frac{\partial^\leftarrow}{\partial p}\frac{\partial^\rightarrow}{\partial q} \right)\right] B \;,
\end{equation}
where the the first-order approximation obviously corresponds to
\begin{eqnarray}
\nonumber
A\star B\approx 
AB+\frac{\hbar'}{2i} \left( \frac{\partial A}{\partial q}\frac{\partial B}{\partial p} -\frac{\partial A}{\partial p}\frac{\partial B}{\partial q} \right)  \\
\label{f2}
=AB+\frac{\hbar'}{2} \left( \frac{\partial A}{\partial a}\frac{\partial B}{\partial a^*} -\frac{\partial A}{\partial a^*}\frac{\partial B}{\partial a} \right)  \;.
\end{eqnarray}
Applying Eq.~(\ref{f2}) to Eq.~(\ref{a3}) we obtain
\begin{eqnarray}
\nonumber
{\cal L}_\gamma(f)=\frac{\gamma}{2\hbar'}(a^*\star a\star f - 2a\star f\star a^* + f\star a^*\star a) \\
\label{f3}
\approx\gamma f +\frac{\gamma}{2}\left( a\frac{\partial f}{\partial a} + a^*\frac{\partial f}{\partial a^*} \right) \;.
\end{eqnarray}

\section{Appendix B}

Let us show that the quantum counterpart of the diffusion term in Eq.~(\ref{c4}) is given by the sum of two Lindblad operators,
\begin{equation}
\label{d1}
\widehat{{\cal D}}({\cal R})=\widehat{{\cal L}}_1({\cal R}) + \widehat{{\cal L}}_2({\cal R}) \;,
\end{equation}
where
\begin{eqnarray}
\label{d2a}
\widehat{{\cal L}}_1({\cal R})
=\frac{D\bar{n}}{2}(\hat{a}\hat{a}^\dagger{\cal R}-2\hat{a}^\dagger{\cal R}\hat{a}  + {\cal R}\hat{a}\hat{a}^\dagger) \;,\quad  \\
\label{d2b}
\widehat{{\cal L}}_2({\cal R})=
\frac{D\bar{n}}{2}(\hat{a}^\dagger\hat{a}{\cal R}-2\hat{a}{\cal R}\hat{a}^\dagger  + {\cal R}\hat{a}^\dagger\hat{a}) \;.
\end{eqnarray}
First of all we note that in the first-order approximation the operators (\ref{d2a}) and (\ref{d2b}) have the same form (see Appendix A) but a different sign. Thus,  $\widehat{{\cal D}}({\cal R})$ does not vanish only in the second-order approximation. To calculate the second-oder Wigner-Weyl image of $\widehat{{\cal D}}({\cal R})$ we rewrite Eq.~(\ref{d1}) in the following form 
\begin{equation}
\label{d3}
\widehat{{\cal D}}({\cal R})=\frac{D\bar{n}}{2}\left( [\hat{a},[\hat{a}^\dagger,{\cal R}]] + [\hat{a}^\dagger,[\hat{a},{\cal R}]] \right)  \;.
\end{equation}
Using the standard correspondence relation where the commutator corresponds to the Poisson brackets, we obtain from Eq.~(\ref{d3})
\begin{equation}
\label{d4}
{\cal D}(f)= \frac{D}{2}\left( \{a,\{a^*,f\}\}+\{a^*,\{a,f\}\} \right) = D \frac{\partial^2 f}{\partial a \partial a^*} \;.
\end{equation}

Next we discuss limitations of Eq.~(\ref{d1}) and Eq.~(\ref{d4}).  As it is easy to prove, the diffusion term leads to unbounded growth of the oscillator action as $I=Dt$. This drawback of the model can be eliminated by introducing an effective friction
\begin{equation}
\label{d5}
{\cal L}_{eff}(f)=D f +\frac{D}{2}\left( a\frac{\partial f}{\partial a} + a^*\frac{\partial f}{\partial a^*} \right) \;.
\end{equation}
Notice that this additional term naturally appears in the equation on the classical distribution function if we begin with the bosonic relaxation operator of the standard form 
\begin{eqnarray}
\nonumber
\widehat{{\cal L}}({\cal R})=\frac{D}{2}\left[
\bar{n}  (\hat{a}\hat{a}^\dagger{\cal R}-2\hat{a}^\dagger{\cal R}\hat{a}  + {\cal R}\hat{a}\hat{a}^\dagger)  \right. \\
\label{d6}
\left. + (\bar{n}+1) (\hat{a}^\dagger\hat{a}{\cal R}-2\hat{a}{\cal R}\hat{a}^\dagger  + {\cal R}\hat{a}^\dagger\hat{a}) \right]  \;,
\end{eqnarray}
where the parameter $D$ is usually referred to as the decay constant. It should be understood, however, that the physical meaning of this parameter is the diffusion constant rather than a decay constant. This becomes particularly clear in the case of the induced decay with the rate $\gamma \gg D$, where one can neglect the `internal friction'  (\ref{d5}) in comparison with the `external  friction' (\ref{c1}). Of course, devision of the relaxation process into diffusion and friction is valid only in the pseudoclassical limit $\bar{n}\gg 1$. If $\bar{n}\sim 1$ one typically speaks about the decoherence and relaxation. With respect to the Bose-Hubbard reservoir with $\bar{n}\sim 1$ the validity of the master equation with the relaxation term (\ref{d6}) was discussed in Ref.~\cite{108}, with the main conclusion  that the necessary and sufficient condition of its validity is the condition of Quantum Chaos in the Bose-Hubbard model.

\end{document}